\documentclass[apj]{emulateapj}
\bibpunct{(}{)}{;}{a}{}{,}

\begin{document}
\title{Hard X-ray and ultraviolet emission during the 2011 June 7 solar flare}

\author{A. R. Inglis\altaffilmark{1} and H. R. Gilbert}
\affil{Solar Physics Laboratory, Heliophysics Science Division, NASA Goddard Space Flight Center, Greenbelt, MD, 20771, USA}

\altaffiltext{1}{Physics Department, The Catholic University of America, Washington, DC, 20664, USA}

\begin{abstract}

The relationship between X-ray and UV emission during flares, particularly in the context of quasi-periodic pulsations, remains unclear. To address this, we study the impulsive X-ray and UV emission during the eruptive flare of 2011 June 7 utilising X-ray imaging from RHESSI and UV 1700$\AA$ imaging from SDO/AIA. This event is associated with quasi-periodic pulsations in X-ray and possibly UV emission, as well as substantial parallel and perpendicular motion of the hard X-ray footpoints. 

The motion of the footpoints parallel to the flare ribbons is unusual; it is shown to reverse direction on at least two occasions. However, there is no associated short-timescale motion of the UV bright regions. Over the same time interval, the footpoints also gradually move apart at $v \approx $ 12 km/s, consistent with the gradual outward expansion of the UV ribbons and the standard flare model. Additionally, we find that the locations of the brightest X-ray and UV regions are different, particularly during the early portion of the flare impulsive phase, despite their integrated emission being strongly correlated in time. Correlation analysis of measured flare properties, such as the footpoint separation, flare shear, photospheric magnetic field and coronal reconnection rate, reveals that - in the impulsive phase - the 25 - 50 keV hard X-ray flux is only weakly correlated with these properties, in contrast to previous studies.

We characterise this event in terms of long-term behaviour, where the X-ray nonthermal, thermal, and UV emission sources appear temporally and spatially consistent, and short-term behaviour, where the emission sources are inconsistent and quasi-periodic pulsations are a dominant feature requiring explanation. We suggest that the short timescale behaviour of hard X-ray footpoints, and the nature of the observed quasi-periodic pulsations, is determined by fundamental, as-yet unobserved properties of the reconnection region and particle acceleration sites. This presents a challenge for current three-dimensional flare reconnection models.

\end{abstract}

\maketitle

\section{Introduction}
\label{intro}
 
A complete understanding of the energy release process that causes solar flares remains elusive. In the standard CSHKP model \citep{1964NASSP..50..451C, 1974SoPh...34..323H, 1966Natur.211..695S, 1976SoPh...50...85K} the advent of magnetic reconnection in the solar corona releases substantial energy, generating emission over a wide spectrum, from radio waves and microwaves to X-rays and gamma-rays. However, the exact circumstances that lead to a flare occuring remain the subject of active study.

One of the main characteristics of flares is the acceleration of particles, which propagate along field lines to the chromosphere and interact according to the thick target model \citep{1971SoPh...18..489B}, generating bremsstrahlung radiation which is observed in the hard X-ray regime. Observations of hard X-ray emission from Yohkoh and the Reuven Ramaty High Energy Solar Spectroscopic Imager \citep[RHESSI;][]{Lin2002} have shown that this emission generally takes the form of two footpoints separated by the magnetic neutral line. Furthermore, many studies have shown that these footpoints often move significantly over time during the flare \citep[e.g.][]{2002SoPh..210..307F, 2003ApJ...595L.103K, 2005ApJ...625L.143G, 2005ApJ...630..561B, 2005AdSpR..35.1707K, 2009ApJ...693..132Y, 2009ApJ...693..847L, 2012ApJ...748..139I}. This motion can be parallel to the magnetic neutral line, perpendicular, or a combination of the two \citep[e.g.][classified the motion of a number of events]{2009ApJ...693..132Y}, with typical velocities in the range 20 - 200 km s$^{-1}$ \citep{2003ApJ...595L.103K, 2005AdSpR..35.1707K}. Similar footpoint motions are observed at other wavelengths, such as white light and ultraviolet \citep[e.g.][]{2006ApJ...641.1217C, 2009A&A...493..241F}. These motions complicate the understanding of the flare process, as they imply either the movement of the flare reconnection site or the existence of many distinct reconnection events during a flare. For a recent review of solar flare X-ray observations, see \citet{2011SSRv..159..107H}.

The ultraviolet (UV) emission from flares often takes the form of two elongated ribbons aligned approximately parallel to the neutral line \citep[e.g.][]{2006ApJ...641.1197S}. This ribbon emission is thought to originate from the solar chromosphere, and the exact ribbon morphology is dependent on the magnetic configuration of the active region \citep[e.g.][]{1997A&A...325..305D}. The emission mechanism responsible for producing UV and optical flare emission is not yet well-understood. One scenario is that electron beams directly heat the lower solar atmosphere \citep[e.g.][]{1970SoPh...15..176N, 1972SoPh...24..414H}, however it has been shown that only very energetic electrons could penetrate to the lower atmosphere and contribute to direct heating. Alternatively, it has been proposed that electron beams deposit their energy in the upper chromosphere, and that this energy is subsequently transferred to the lower chromosphere and photosphere via `radiative backwarming' \citep[e.g.][]{1990ApJ...350..463M, 2003A&A...403.1151D}. For a discussion of the proposed mechanisms for optical and UV emission during flares see \citet{2006ApJ...641.1210X}; see also \citet{2010ApJ...725..319Q, 2013arXiv1305.6899Q} for recent studies of the relationship between UV emission and hard X-ray sources. A full review of the observational properties of flares may be found in \citet{2011SSRv..159...19F}.

Another common characteristic of flares is the presence of quasi-periodic pulsations (QPP) in the flare lightcurves. This phenomenon has been observed for several decades \citep[e.g.][]{1969ApJ...155L.117P, 1970SoPh...13..420C}, and recent observations have provided examples of QPP throughout the electromagnetic spectrum \citep[e.g.][]{2001ApJ...562L.103A,2003ApJ...588.1163G,2005ApJ...625L.143G,2005A&A...440L..59F,2005A&A...439..727M, 2008SoPh..247...77L, 2008A&A...487.1147I, 2009SoPh..258...69Z, 2010ApJ...708L..47N, 2011A&A...525A.112R, 2012ApJ...748..139I}. The nature of QPP remains debated. One possibility is that they are a manifestation of magnetohydrodynamic (MHD) wave processes excited in flare sites, which allows for the estimation of solar plasma parameters via coronal seismology. Alternatively, QPP may be a fundamental signature of the energy release process during flares, which may occur in a bursty fashion \citep[e.g.][]{2006ApJ...642.1177L, 2011ApJ...730...90G, 2012A&A...548A..98M}, see \citet{2009SSRv..149..119N} for a full review. In either case, explanations of QPP must be reconciled with the observed sub-sonic motions of X-ray footpoints and  current 3-dimensional flare models.

In this work we present observations of the hard X-ray and UV emission during the 2011 June 7 flare \citep[IAU standard: SOL2011-06-07T06:24:00L045C112, see][for more details]{2010SoPh..263....1L}, an event of GOES-class M2.6 which originated from active region AR 11226 at 06:24 UT. This event featured remarkable motions of the hard X-ray footpoints during the impulsive phase, and also exhibited QPP in the X-ray lightcurves. The UV emission takes the form of two elongated ribbons, in which evidence of QPP was also observed. In addition, this event was a spectacular eruption, featuring the ejection of a large amount of prominence material, much of which failed to escape from the Sun's gravitational influence and returned to the solar surface. This ejected material has been previously studied by \citet{2012A&A...540L..10I}, \citet{2013ApJ...764..165W} and \citet{Gilbert2013}. The flare was also shown by \citet{2012ApJ...746...13L} to be the source of a globally propagating EUV wave, and was shown by \citet{2013arXiv1304.3749F} to be the source of substantial post-flare gamma-ray emission.

\section{The event: 2011 June 7}
\subsection{X-ray lightcurves}
\label{sec_lightcurves}

\begin{figure}[h]
\begin{center}
\includegraphics[width=8cm]{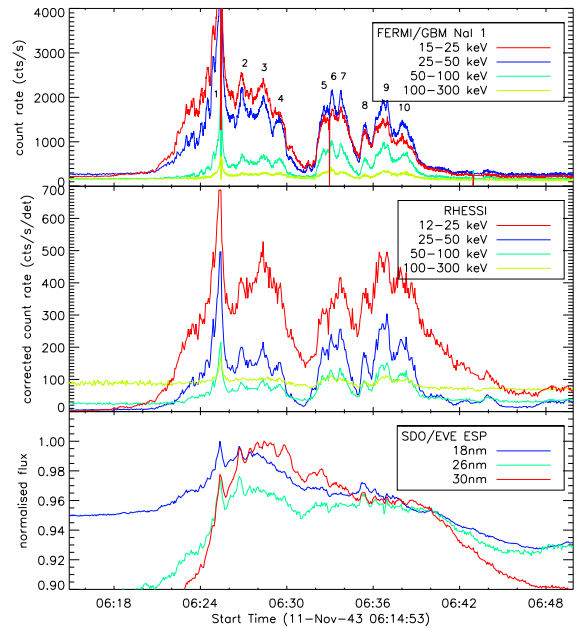}
\caption{Lightcurve of the 2011 June 7 flare as observed by various instruments. Top panel: X-ray count rates observed by Fermi/GBM in a range of energies. Center panel: X-ray count rates observed by RHESSI over a range of energies. Bottom panel: Normalised EVE/ESP fluxes from three of the observing channels. The numerals in the top panel denote clearly identifiable pulses in the hard X-ray emission. }
\label{lightcurve}
\end{center}
\end{figure}

This flare was observed in the X-ray regime by both RHESSI and the Fermi Gamma-Ray Burst Monitor \citep[GBM;][]{2009ApJ...702..791M}. Figure \ref{lightcurve} shows the lightcurves from both instruments over a range of energies. Here, data from one of the more sunward-pointed GBM detectors is shown (detector 1 in this case), while the RHESSI data are averaged over all nine of its detectors. As expected, a close correlation between these two datasets is evident, as are a number of lightcurve peaks. The differences in count rates observed by Fermi/GBM relative to RHESSI are partly due to the angle between the chosen GBM detector and the Sun, and also due to the application of attenuators on board the RHESSI spacecraft, which automatically cover its detectors during periods of strong activity, reducing count rates. An additional consequence of this is that the GBM detectors, lacking attenuators, are more susceptible to pulse-pileup effects at low energies during periods of high count rates, i.e. strong flares. The bottom panel of Figure \ref{lightcurve} shows lightcurves obtained from the EVE/ESP \citep{2012SoPh..275..115W} instrument on board the Solar Dynamics Observatory \citep[SDO;][]{2012SoPh..275....3P}, where the flux has been re-binned to a 2 s cadence to improve the signal-to-noise ratio.

In hard X-rays, the strongest emission peak occurs at 06:25:45 UT,  observed in almost all of the energy channels shown in Figure \ref{lightcurve}. This is followed by a period of bursty hard X-ray emission during which several lightcurve peaks are observed, continuing until approximately 06:41 UT. For reference we number these peaks 1 - 10 as shown in Figure \ref{lightcurve}. The appearance of such pulsations is important; it suggests either the dynamic and possibly periodic variation of parameters in the flare reconnection or particle acceleration sites, or the triggering of a secondary mechanism within the flare which is capable of modulating broadband flare emission \citep{2009SSRv..149..119N}. Moreover, the relationship between the properties of X-ray pulsations and the behaviour of X-ray footpoints and other observable flare parameters remains unclear, despite recent studies \citep[e.g.][]{2011ApJ...730L..27N, 2012ApJ...748..139I}. Exploring this relationship provides one of the main motivations of the present work.

Previous studies have suggested there may be strong correlations between the EUV and hard X-ray signatures of QPP during flares \citep{2012ApJ...749L..16D}. In the EUV regime as observed by EVE, there appear to be counterparts to some of the hard X-ray peaks, particularly peaks 1 - 4. At later times the situation becomes less clear; the thermal emission becomes complex and decorrelated from the hard X-ray emission. This decorrelation may be due to the ejection of prominence material and secondary heating of the active region during this event, which may account for a substantial fraction of the spatially-integrated emission observed by EVE. 

\subsection{X-ray footpoint motions}
\label{sec_footpoints}

In order to investigate the motion of the hard X-ray footpoints, we follow the approach of \citet{2012ApJ...748..139I}, who analysed footpoint progression in three flares observed by RHESSI. Here, we summarize the key points of the method.

\begin{figure*}
\begin{center}
\includegraphics[width=8cm,bb = 180 0 640 425]{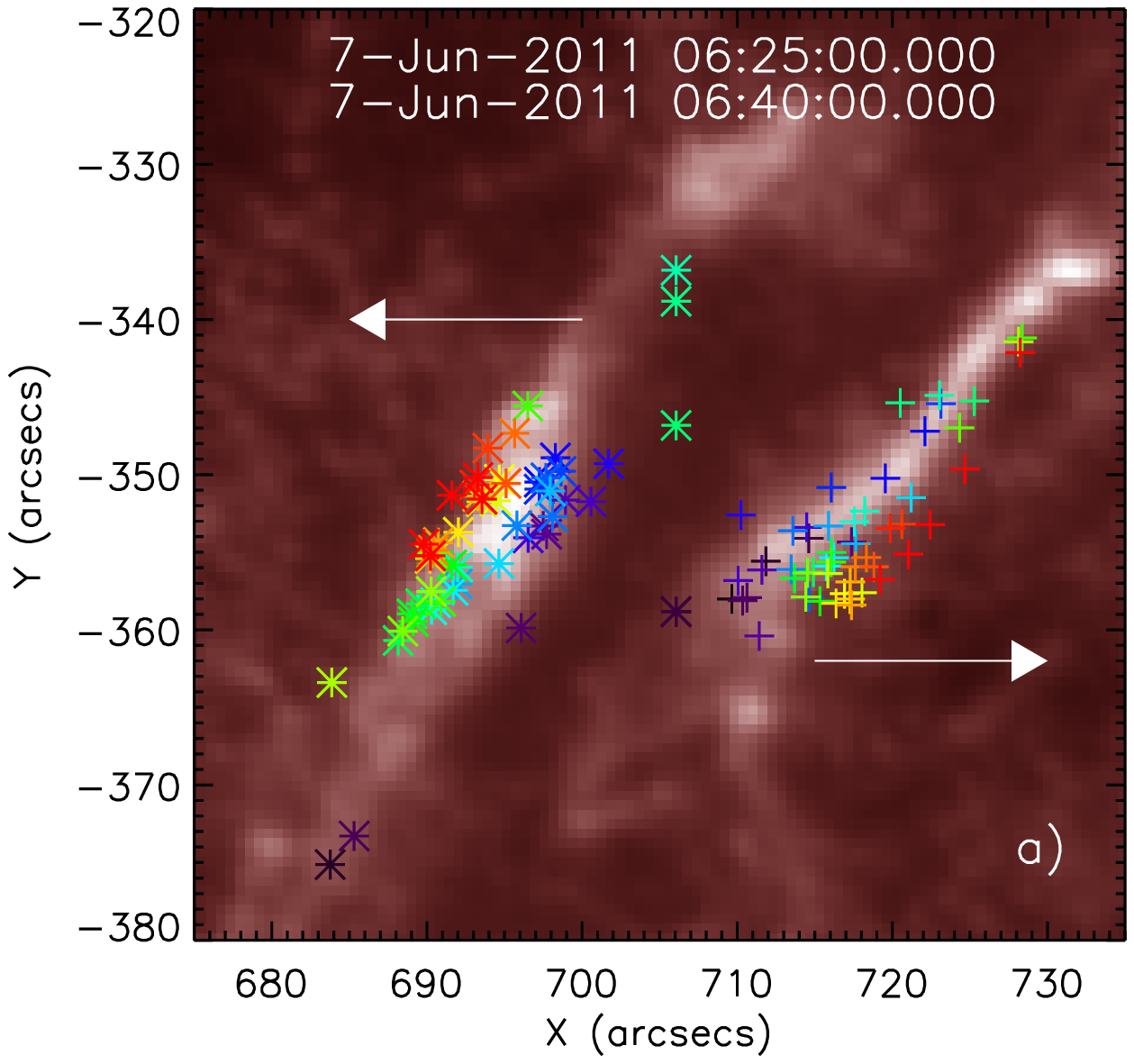}
\includegraphics[width=0.58cm,bb = 0 -55 56 350]{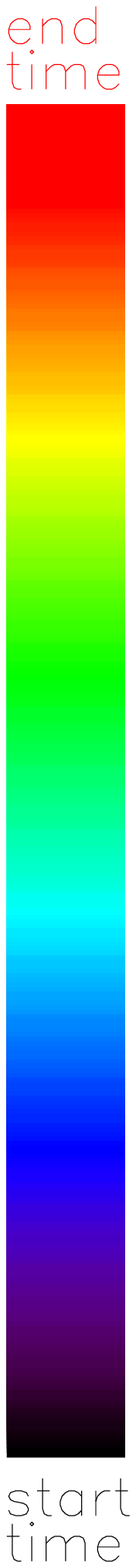}
\includegraphics[width=8cm,bb = 180 0 640 425]{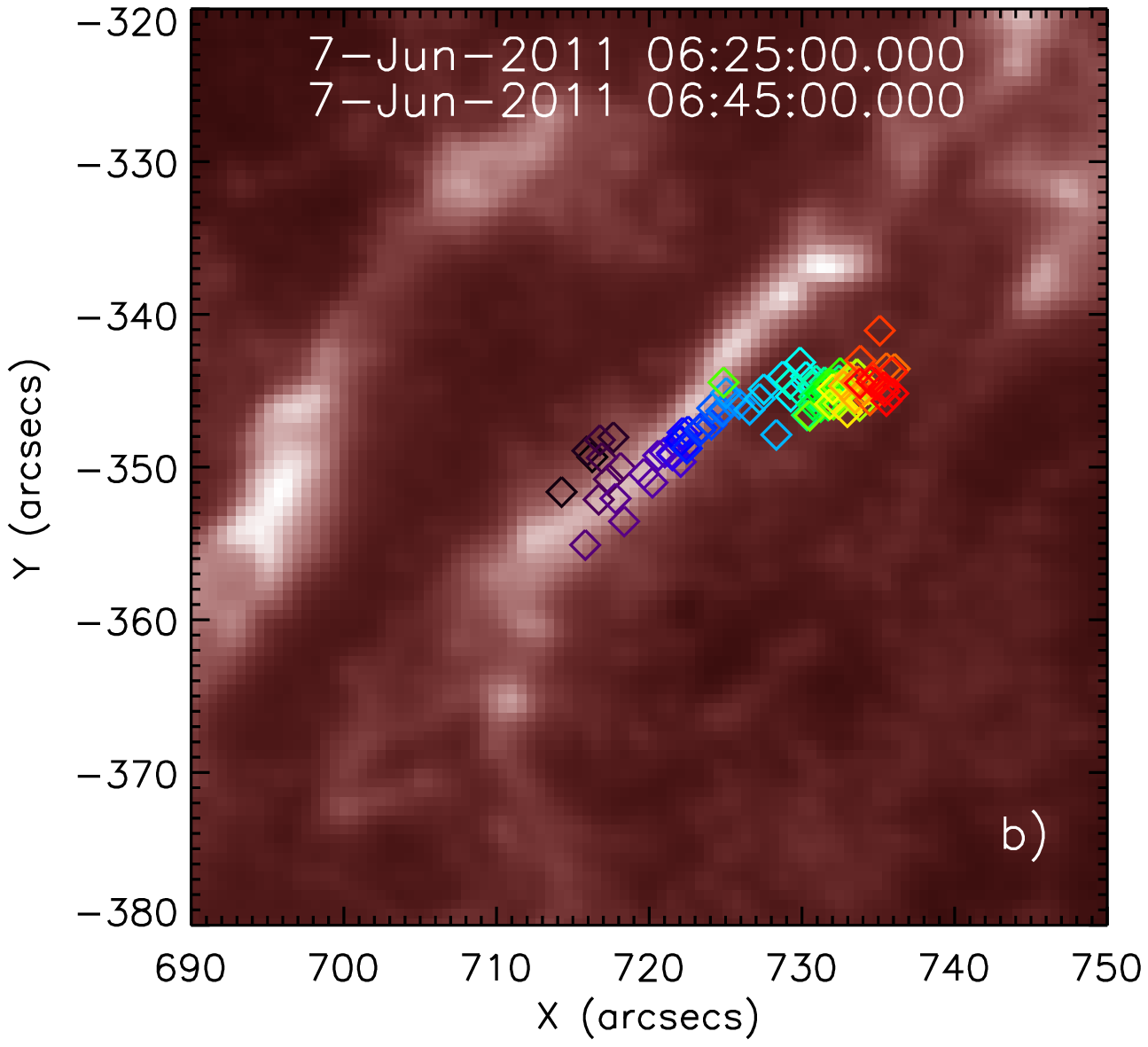}
\includegraphics[width=18cm]{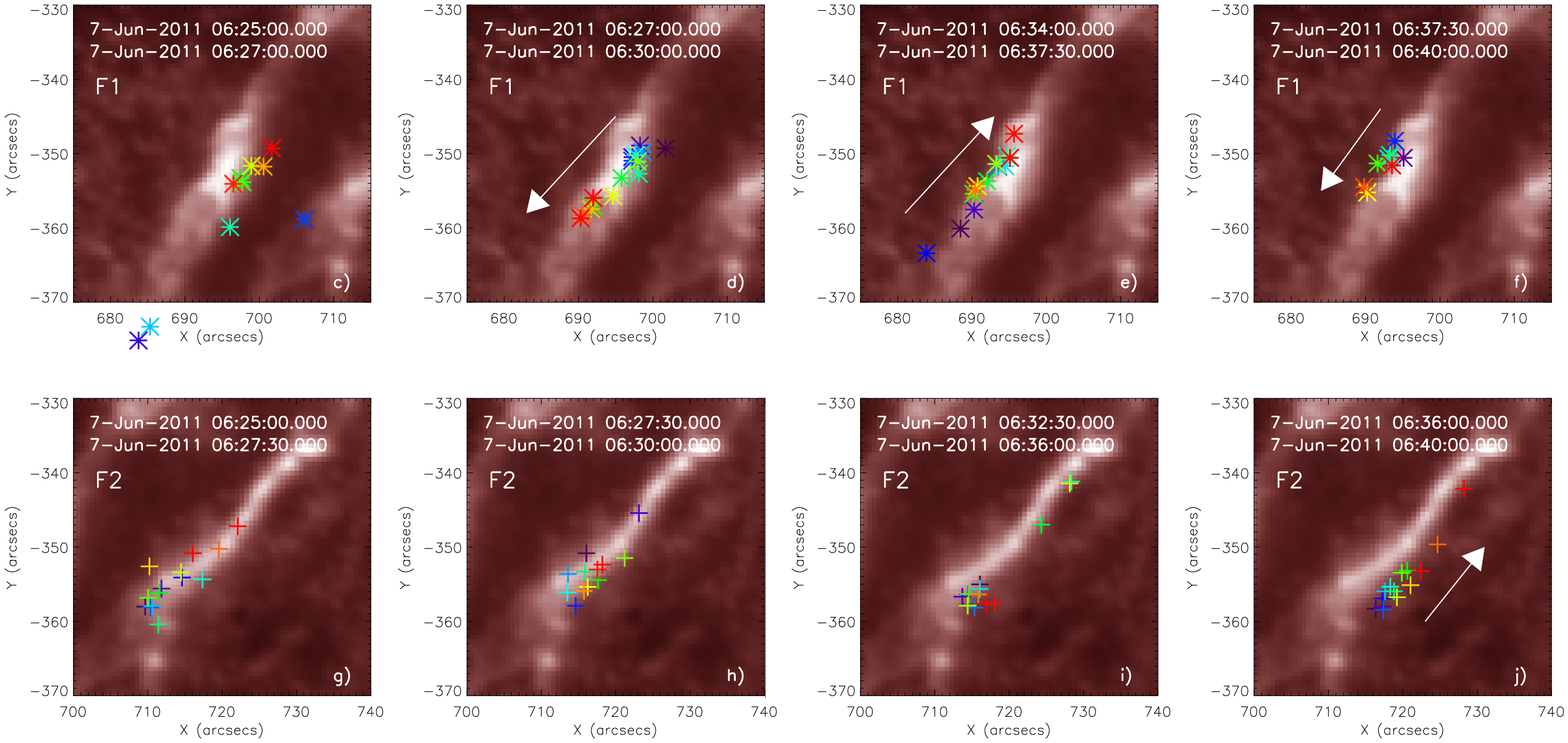}
\caption{a) Hard X-ray footpoint locations in the 25-50 keV range as a function of time during the 2011 June 07 eruption, as observed by RHESSI. The footpoint locations are denoted by the coloured markers, while the background image is taken by AIA at 06:31:19 UT in the 1700$\AA$ range. In each panel the footpoint colours progress from blue (start time) to red (end time). b) Locations of the thermal X-ray source in the 6-12 keV energy band as a function of time. c)-f): These four panels show the movement of the left-hand footpoint (F1) at different time intervals as follows: c) 06:25 - 06:27 UT, d) 06:27 - 06:30 UT, e) 06:34 - 06:37:30, and f) 06:37:30 - 06:40 UT. Panels g)-j): show the movement of the right-hand footpoint (F2) at different time intervals as follows: g) 06:25 - 06:27:30 UT, h) 06:27:30 - 06:30 UT, i) 06:32:30 - 06:36 UT, j) 06:36 - 06:40 UT. Note that the time intervals differ slightly between the two footpoints.  }
\label{footpoint_locations}
\end{center}
\end{figure*}

Firstly, a sequence of RHESSI images is generated using the CLEAN algorithm \citep{2002SoPh..210...61H}. For the 2011 June 7 event, the 25 - 50 keV energy range is chosen, corresponding to the non-thermal X-ray regime for this event, and the images are reconstructed using a 12 s cadence. Each image frame is divided into two sections, with the divider running between the two flare ribbons, approximately along the neutral line. At this stage, image frames for which the flux in either section of the frame falls below a threshold of 10 counts s$^{-1}$ cm$^{-2}$ are removed. For the remaining images, the point of peak emission is located in each image section. Taking peak flux location instead of flux centroids leads to increased scatter in the measured footpoint locations. However, as mentioned by e.g. \citet{2002SoPh..210..307F, 2009ApJ...698.2131D}, in strong flares centroid locations can often be adversely affected by background X-ray emission in an image, leading to systematic shifts in estimated footpoint location.

A visual examination of the RHESSI images reveals that, although the majority of image frames show two compact X-ray sources, in some image frames extended or multiple sources are present. This may be partly due to the relatively long 12 s time integration, which may result in two temporally separate sources appearing in the same image. Hence, for these frames the method described above can be considered as an approximation which provides the location of the brightest hard X-ray source in each image section.

The footpoint locations are shown in Figure \ref{footpoint_locations}. Panel a) shows the motion of the footpoints over the entire time duration of the flare in hard X-rays, between 06:23 - 06:41 UT.  Beyond 06:41 UT the flare impulsive phase ends and the 25-50 keV emission falls below our established imaging threshold for reliably reconstructing RHESSI images \citep[see][for a thorough discussion of RHESSI imaging techniques and desired conditions]{2009ApJ...698.2131D}. The footpoint colours progress in time from blue (start time) to red (end time). 

\begin{figure*}
\begin{center}
\includegraphics[width=16.0cm]{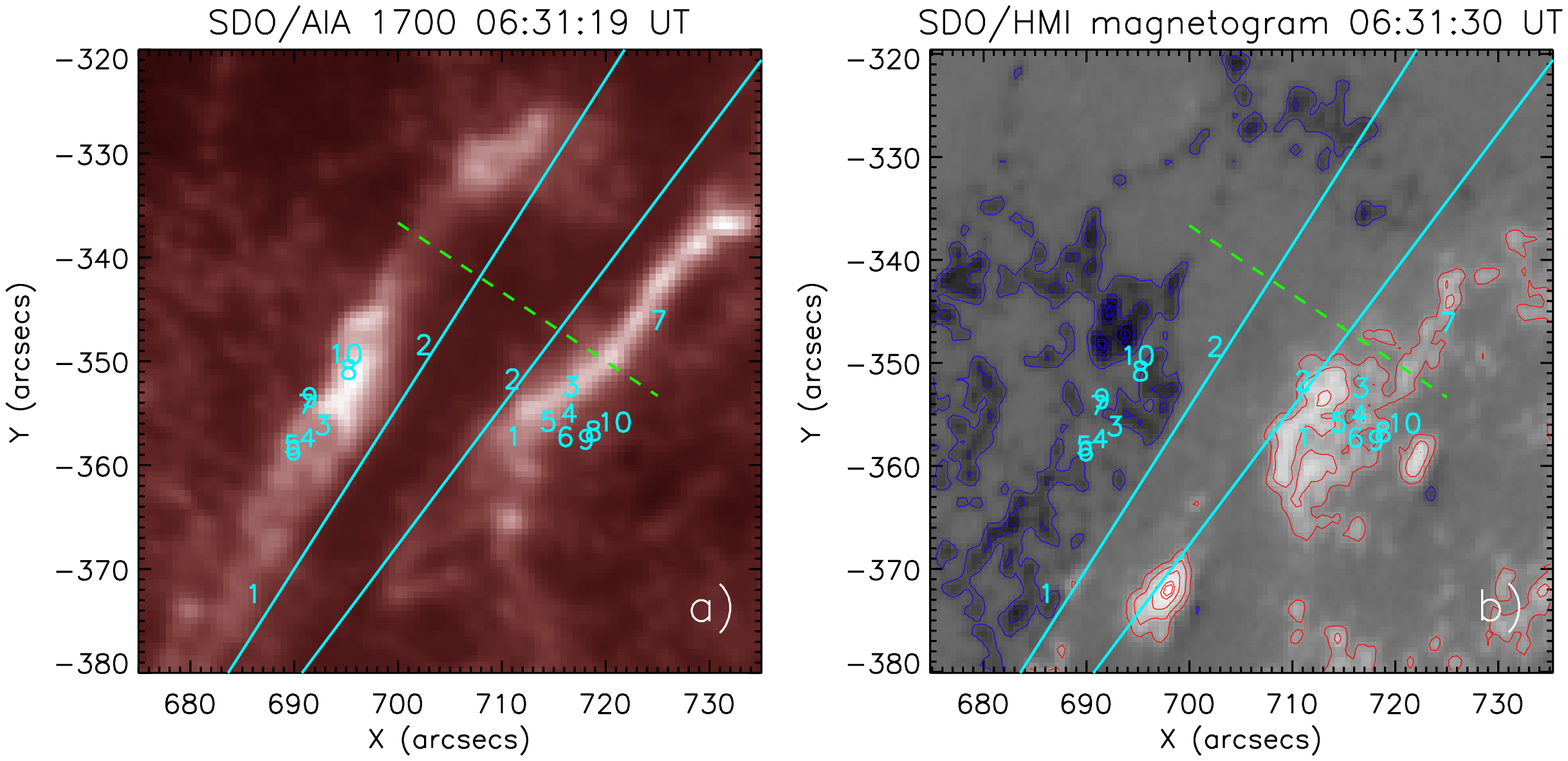}
\includegraphics[width=17.0cm]{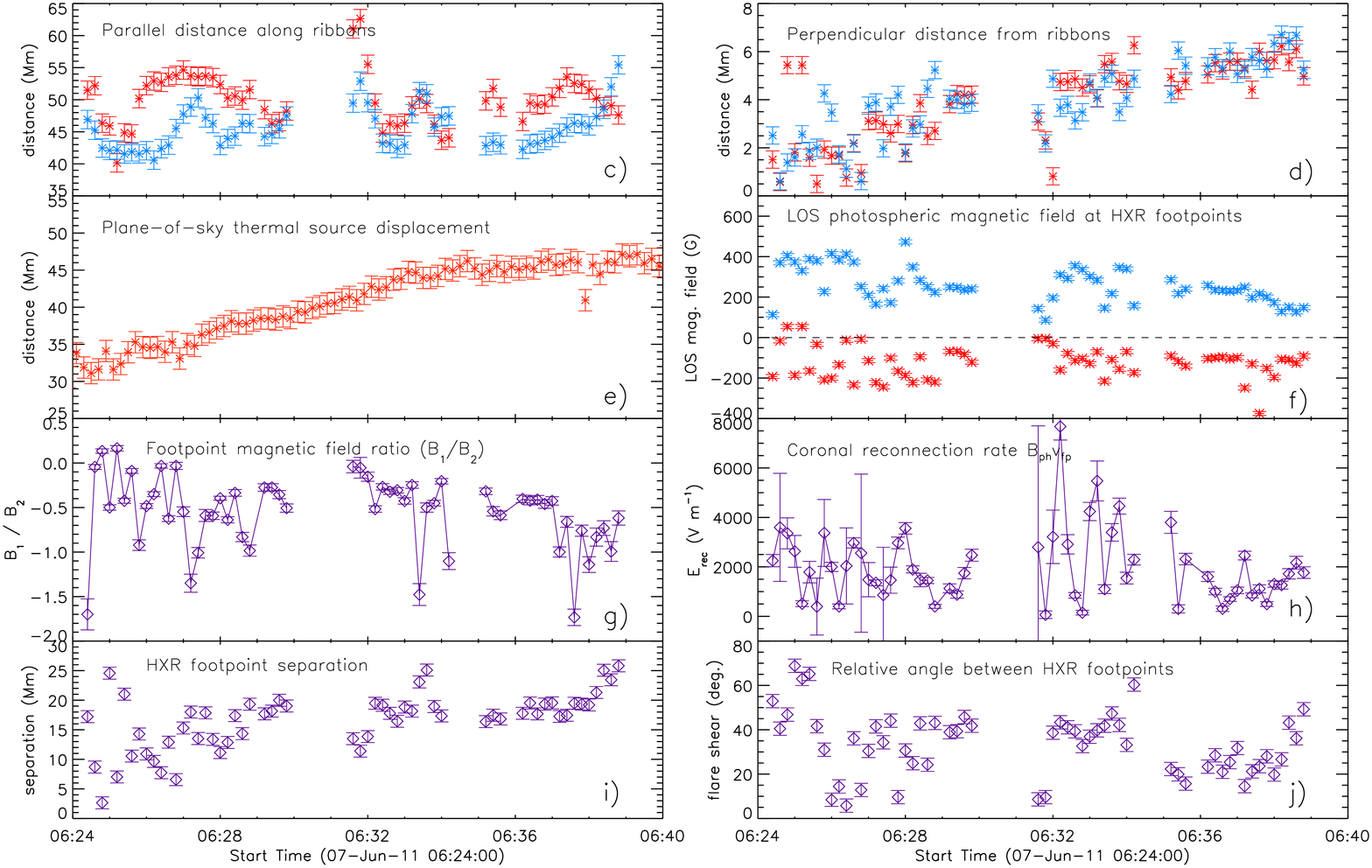}
\caption{a) The locations of the hard X-ray footpoints during the ten peaks identified in Figure \ref{lightcurve}. The two cyan lines represent an approximation of the ribbon locations at the onset of the flare (06:24 UT). The background is an AIA 1700$\AA$ image illustrating the flare ribbons. b) Same X-ray footpoint locations as a), overlayed on a SDO/HMI line-of-sight magnetogram. The red contours highlight strong positive polarity regions, the blue contours highlight strong negative regions. c)-j) Various flare properties as a function of time. c) Movement of sources F1 (red) and F2 (blue), where the distance of each footpoint is calculated from an arbitrary start point. The start points are located to the solar southeast, directly along the ribbon lines illustrated in panel a). d) Distance of F1 and F2 from their respective ribbon lines, as indicated in panel a). e) Plane-of-sky motion of the thermal 6-12 keV source from an arbitrary starting coordinate located to the solar southeast. f) Line-of-sight photospheric magnetic field beneath F1 (red) and F2 (blue). g) Ratio of the photospheric magnetic field underneath F1 and F2. h) Coronal reconnection rate $E_{rec} = v_{fp}B_{ph}$. i) Separation of the two footpoints F1 and F2. j) Variation of the flare shear angle between F1 and F2. }
\label{dist_from_line}
\end{center}
\end{figure*}

Additionally, the location of the thermal X-ray emission is found by generating CLEAN images in the 6 - 12 keV energy range, locating the emission peak for each frame. In flares, thermal X-rays are emitted from plasma loops which have been heated due to energy release, often via chromospheric evaporation. Panel b) illustrates the behaviour of the thermal (6-12 keV) X-ray source between 06:23 - 06:45 UT. The source is observed to move smoothly towards solar west as a function of time. Considering the orientation of the flare arcade and the line of sight angle, there appears to be a component of motion along the arcade in addition to an increase in height of the thermal emission.

Let us refer to the left-hand (Eastern) footpoint as F1, and to the right-hand (Western) footpoint as F2. In the remaining panels of Figure \ref{footpoint_locations} the motions of each footpoint have been highlighted at select time intervals for clarity. In order to best highlight their respective motions, these intervals differ slightly for the two footpoints. Panels c) - f) illustrate the behaviour of F1, where the intervals are as follows: c) 06:25 - 06:27 UT, d) 06:27 - 06:30 UT, e) 06:34 - 06:37:30 UT, f) 06:37:30 - 06:40 UT.

Panels g) - j) illustrate the behaviour of F2 during the following intervals: g) 06:25-06:27:30 UT, h) 06:27:30 - 06:30 UT, i) 06:32:30 - 06:36 UT, j) 06:36 - 06:40 UT.

Initially (Figure \ref{footpoint_locations}c) the motion of F1 is scattered, but shows some evidence of motion to the solar northwest. In Figure \ref{footpoint_locations}d, F1 appears to move generally towards the solar southeast, while in Figure \ref{footpoint_locations}e the situation is reversed, with F1 moving towards the solar northwest. In both time intervals the distance covered is approximately 13 Mm along the flare ribbon. The motion is reversed for a second time between 06:37 - 06:41 UT (Figure \ref{footpoint_locations}f), where the footpoint moves towards solar southeast once more, this time covering a smaller distance of $\approx$ 8 Mm.

From this we find that the mean velocity of F1 between 06:27 - 06:31 UT is $v_{\parallel}$ $\approx$ 55 km/s, while it is $v_{\parallel}$ $\approx$ 36 km/s between 06:31 - 06:37 UT and $v_{\parallel}$ $\approx$ 33 km/s between 06:37 - 06:41. These values are in agreement with previous observations of footpoint velocities along ribbons \citep[e.g.][]{2003ApJ...595L.103K, 2005ApJ...625L.143G, 2012ApJ...748..139I}.

The behaviour of F2 is less clear. In panel \ref{footpoint_locations}g there is no clear trend of footpoint motion; instead the locations of the footpoints are distributed randomly along the flare ribbon. In Figure \ref{footpoint_locations}h there is some evidence that F2 moves towards the solar northwest, covering approximately
10 Mm. Figure \ref{footpoint_locations}i again shows no clear trend of motion, while the clearest motion for F2 is observed in Figure \ref{footpoint_locations}j, where the footpoint moves toward the solar northwest again. Here, the displacement is approximately 10 Mm, with a mean velocity of $v_{\parallel} \approx$ 42 km/s. The out-of-sync behaviour of the two footpoints is not unusual during flares \citep{2011SSRv..159...19F}.

Similarly, using the thermal source locations (discounting sources present prior to the mean impulsive peak at 06:25:45 UT), the plane-of-sky velocity of the thermal emission may be estimated. Between 06:26 - 06:45 UT we find that the plane of sky displacement is $\approx$ 15 Mm (see Figure \ref{dist_from_line}), leading to a mean plane-of sky velocity of $\approx$ 13 km/s.

\subsection{Flare arcade and magnetic field measurements}

In order to further investigate the properties of this flare, it is useful at to define a number of reference points and parameters. Firstly, we can approximate the flare ribbons seen in the 1700$\AA$ images with two near-parallel lines as shown in Figure \ref{dist_from_line}a. The neutral line for this event runs almost parallel to and in-between these defined ribbon lines. It is also convenient at this point to introduce the concept of ``flare shear'' as defined by \citet{2007ApJ...660..893J}, which is the angle formed by a line connecting two flare footpoints and a line perpendicular to the neutral line. This is distinct from the magnetic shear, which is estimated from vector magnetogram data. Here, the line of zero flare shear is estimated and illustrated by the dashed green line in Figure \ref{dist_from_line}, where the angle subtended by this line from the East-West line is 34$^{\circ}$. Using these defined lines as reference points, we investigate the evolution of the footpoint location, separation, and flare shear angle.

Photospheric magnetic field measurements are also available for this flare from the Heliospheric Magnetic Imager (HMI) on board SDO. Figure \ref{dist_from_line}b shows the line-of-sight (LOS) magnetic field strength, with the same reference points overlayed as Figure \ref{dist_from_line}a. The red contours highlight areas of strong positive polarity, while the blue contours highlight strongly negative field strength areas.

From the HMI data, we estimate the magnetic field strength underneath each hard X-ray footpoint as a function of time. In order to account for the uncertainty in the footpoint location, the magnetic field is averaged locally over a 7x7 pixel region centered on each footpoint location, corresponding approximately to the $\pm$ 2'' uncertainty associated with the footpoint locations.

Measurement of the photospheric field also enables an estimate of the coronal reconnection rate $E_{rec} = v_{fp} B_{ph}$ \citep[e.g.][]{2004ApJ...604..900Q, 2004SoPh..222..279F,2005AdSpR..35.1707K, 2007ApJ...654..665T, 2009A&A...493..241F}. Although this simple relation between $E_{rec}$, footpoint velocity and photospheric magnetic field strength should only hold in 2 - 2.5D scenarios \citep{2009A&A...493..241F, 2011SSRv..159...19F}, several studies have shown correlations between this quantity and the flare flux \citep[for example][]{2005AdSpR..35.1707K}.

In Figures \ref{dist_from_line}c-\ref{dist_from_line}j we estimate the parameters described above as a function of time. Figure \ref{dist_from_line}c shows the distance of F1 (red) and F2 (blue) from an arbitrary start point as a function of time. These start points are located directly along the ribbon lines as shown in Figure \ref{dist_from_line}a, where they are chosen to be far to the solar southeast. This clarifies the reversal of the footpoint parallel motion, particularly for F1. For example, between 06:25 - 06:27 UT there is a substantial increase of 10 Mm in F1 distance along the ribbon. This motion is reversed during the interval 06:27 - 06:30 UT. Between 06:30 - 06:32 UT reliable imaging is not available due to the low X-ray count rates at those times. This is followed at 06:32 UT by some discontinuous motion of F1. Subsequently, between 06:33 - 06:37:30 UT the distance to the footpoint increases again, before reversing direction again at 06:37:30 UT. The behaviour of F2 is similar in the 06:33 - 06:37:30 UT interval, although whether there is a systematic trend in direction is debatable. Beyond 06:37:30 UT, F2 exhibits its clearest motion in a single direction, with the distance to the footpoint increasing steadily until the end of the observation interval at 06:40 UT - the opposite motion from that observed in F1.

Meanwhile, Figure \ref{dist_from_line}d shows that, in addition to the motions observed along the flare ribbons, both footpoints move gradually apart, with a displacement of approximately 6 Mm each between 06:24 and 06:40 UT. The figure shows the increasing distance of F1 and F2 from the ribbon lines over time, where the closest approach to the ribbon lines defined in Figure \ref{dist_from_line}a is calculated at each time interval. Hence, some scatter of the points should be expected as F1 and F2 move along the ribbons. From Figure \ref{dist_from_line}d we find that $v \approx $ 6 km/s for footpoint, hence the two footpoints are separating at a relative rate of $v_{\perp} \approx$ 12 km/s. This is interesting to compare with the plane-of-sky motion of the thermal X-ray source (Figure \ref{dist_from_line}e). From context images we can determine that the apparent location of the thermal source close to the west ribbon is purely a line-of-sight effect due to the angle of observation and the location of the active region near the limb. Nevertheless, the thermal source potentially has components of motion both along the flare arcade - particularly in the early phase - and outward or upward. Such motions of the thermal source along the arcade have been observed before \citep{2008ApJ...685L..87L}, but in this event it is difficult to disambiguate the possible directions of motion.

Figures \ref{dist_from_line}f, \ref{dist_from_line}g and \ref{dist_from_line}h show the photospheric magnetic field strength beneath the hard X-ray footpoints, and the derived coronal reconnection rate $E_{rec}$, which is obtained by averaging the measurements of $v_{fp} B_{ph}$ for each footpoint \citep{2004ApJ...604..900Q}. Figure \ref{dist_from_line}f shows that, with the exception of the last few minutes of impulsive emission, the photospheric field associated with F1 is consistently weaker than that associated with F2, with a mean ratio of $\approx$ -0.5.

The separation of the footpoints is plotted as a function of time in Figure \ref{dist_from_line}i, where we measure the trigonometric distance between the locations of F1 and F2 at each time interval. Meanwhile, Figure \ref{dist_from_line}j shows the estimated flare shear angle between the two footpoints, where the shear is estimated as the deviation in degrees from the zero flare shear line shown in Figure \ref{dist_from_line}a \citep{2007ApJ...660..893J}. There is no clear trend in the measurement of the flare shear, except for a sharp decrease in the shear from 06:35 UT onwards. Since it is the trigonometric separation of the hard X-ray footpoints that we measure in Figure \ref{dist_from_line}i, this decrease in flare shear has the apparent effect of counterbalancing the gradual separation of the flare ribbons, the relatively flat separation between 06:32 - 06:38 UT.

\subsection{Ultraviolet ribbons}
\label{uv_ribbons}

\begin{figure}
\begin{center}
\includegraphics[width=10cm,bb=60 10 474 330]{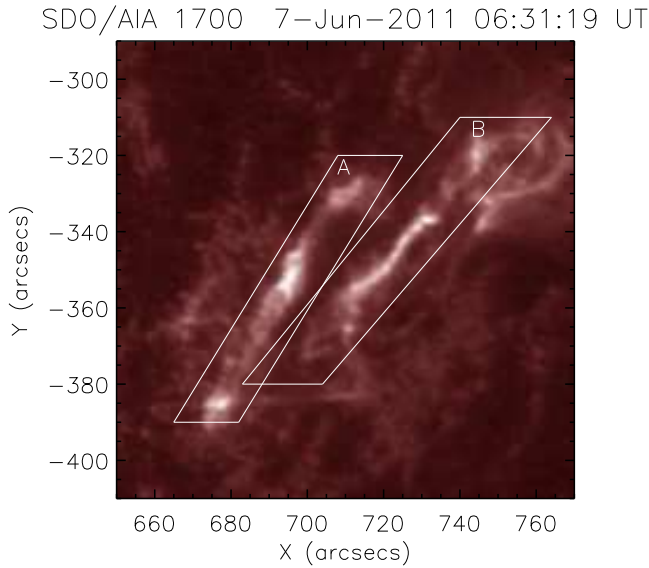}
\caption{SDO/AIA 1700$\AA$ image showing the two ultraviolet ribbons during the 2011 June 7 flare. The two parallelograms denote the user-defined areas of the left-hand (A) and right-hand (B) ribbon. These areas encompass the bright ribbons at all time intervals.}
\label{ribbon_areas}
\end{center}
\end{figure}

\begin{figure*}
\begin{center}
\includegraphics[width=18cm]{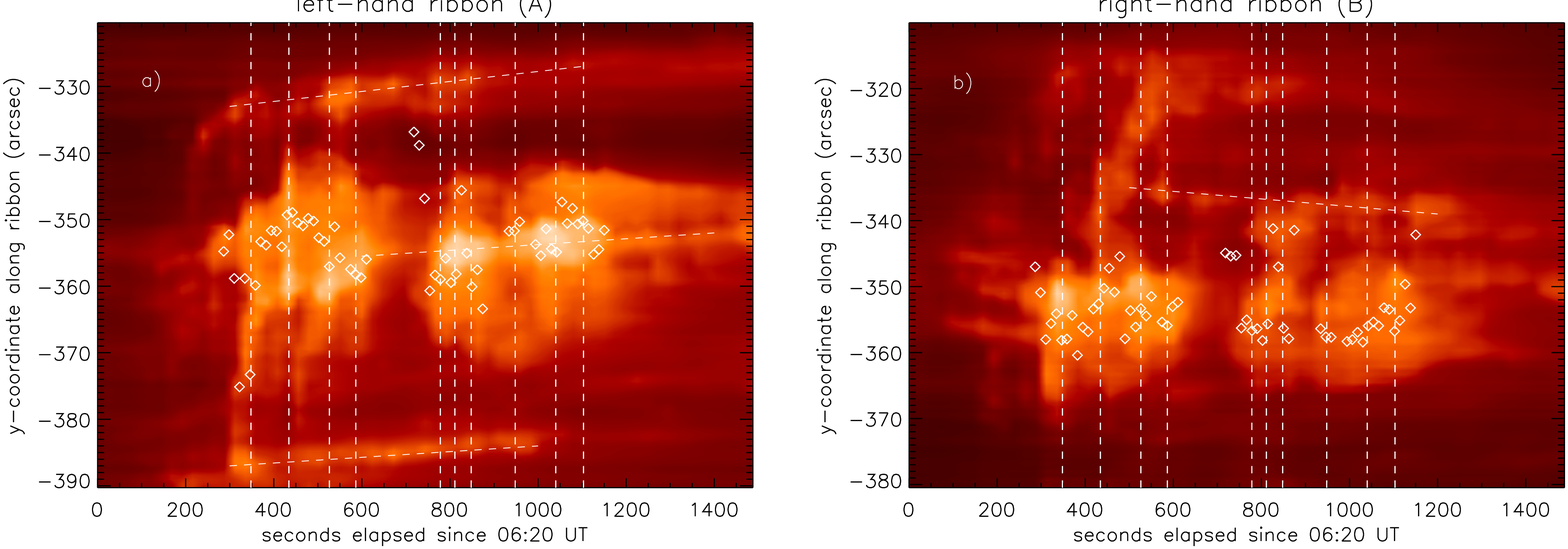}
\includegraphics[width=18cm]{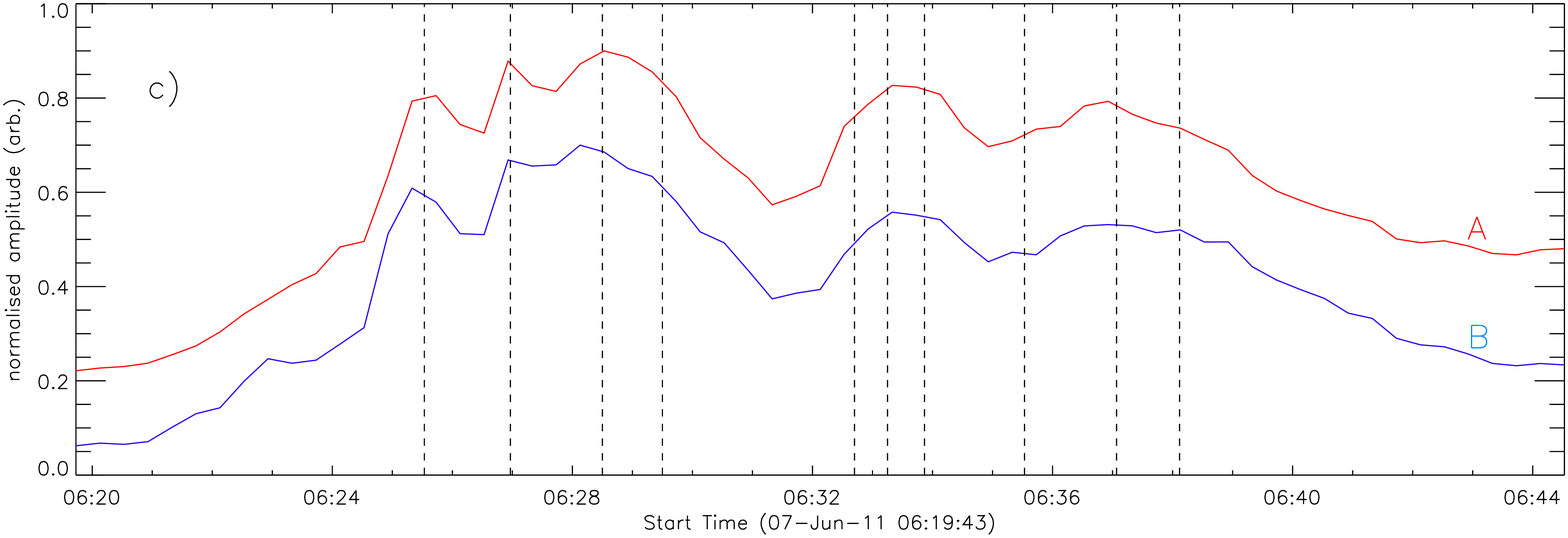}
\caption{\textit{Panels a) and b)}: Time-distance plots of the left-hand and right-hand ribbons, as observed at 1700$\AA$ by AIA. For each ribbon, the x-coordinate has been integrated over to provide the intensity as a function of $y$. The vertical dotted lines illustrate the timing of each pulse observed in the hard X-ray lightcurves by RHESSI. The overlaid crosses in each panel denote the y-position as a function of time of the hard X-ray source associated with the respective ribbon. For image clarity the uncertainties on the hard X-ray source locations are not displayed. \textit{c)}: The total intensity of ribbon A (red) and ribbon B (blue) as a function of time. The vertical dotted lines illustrate the hard X-ray pulse timings.}
\label{time_distance}
\end{center}
\end{figure*}

In this section we examine the flare ribbons observed in the 1700$\AA$ ultraviolet continuum channel by AIA. These ribbons gradually move apart during the flare at a rate of $v \approx$ 12 km/s in tandem with the outward expansion of the hard X-ray footpoints. Hence, in order to capture the ribbon intensity at all image times, two regions are defined on the AIA image which represent the boundaries of each ribbon area (Figure \ref{ribbon_areas}). We define these areas as A and B. Within these areas, we integrate over the solar East-West direction ($x$) in order to find the intensity of the ribbon as a function of the solar North-South direction ($y$). By repeating this process for each image frame, we create the time-distance plots shown in Figure \ref{time_distance}.

It can be seen that the intensity along the ribbons is not uniform, consisting of distinct bright and dark areas in each case. In ribbon A for example, there are three clear brighter areas. The first lies near the southern end of the ribbon, while the second encompasses a substantial part of the ribbon center and the third is found at the northern end of the ribbon. The dashed lines in Figure \ref{time_distance} track the position of these features as a function of time, showing that they remain consistently located throughout the flare. Each feature displays a displacement of a few arcseconds towards the ribbon's northern end. This behaviour can be explained by the fact that the ribbon motion is not purely in the solar East-West direction which we have integrated over; there is an additional component in the North-South direction, as the ribbons expand outward from the magnetic neutral line.

In Figure \ref{time_distance} the timing of each pulse observed in hard X-rays (see Figure \ref{lightcurve}) is denoted by the vertical dotted lines in each panel. Figure \ref{time_distance}c shows the total intensity of each ribbon as a function of time. From this it can be seen that, generally speaking, the hard X-ray peaks correspond to brightenings in the UV ribbon intensity, indicating a link between the UV brightenings and particle acceleration. However, as has been noted, this link remains unclear, and UV emission may also be generated by other effects \citep[e.g.][]{2006ApJ...641.1210X, 2010ApJ...725..319Q}. Peaks 1-3 are particularly well correlated with counterparts in the UV emission. The valley in X-ray emission seen between 06:30 - 06:32 UT - which separates peaks 4 and 5 - is also clearly replicated in the UV emission intensity of each ribbon.

Overlaid on Figure \ref{time_distance}a and Figure \ref{time_distance}b are the positions of the hard X-ray source associated with each ribbon. This illustrates that for both ribbon A and B the hard X-ray footpoints are confined to the central, bright ribbon areas. However, the hard X-ray footpoint locations are not in general cospatial with the brightest points in the UV ribbons. For peaks 1-4 there is a clear offset for both ribbons between the hard X-ray and UV peak locations. For the remaining peaks the situation is less clear. In ribbon A the offset continues for peaks 5-7, whereas peaks 8-10 are approximately cospatial. In ribbon B, the UV ribbon is more diffuse, making the comparison of peak emission locations difficult.

\begin{figure}
\begin{center}
\includegraphics[width=8.5cm]{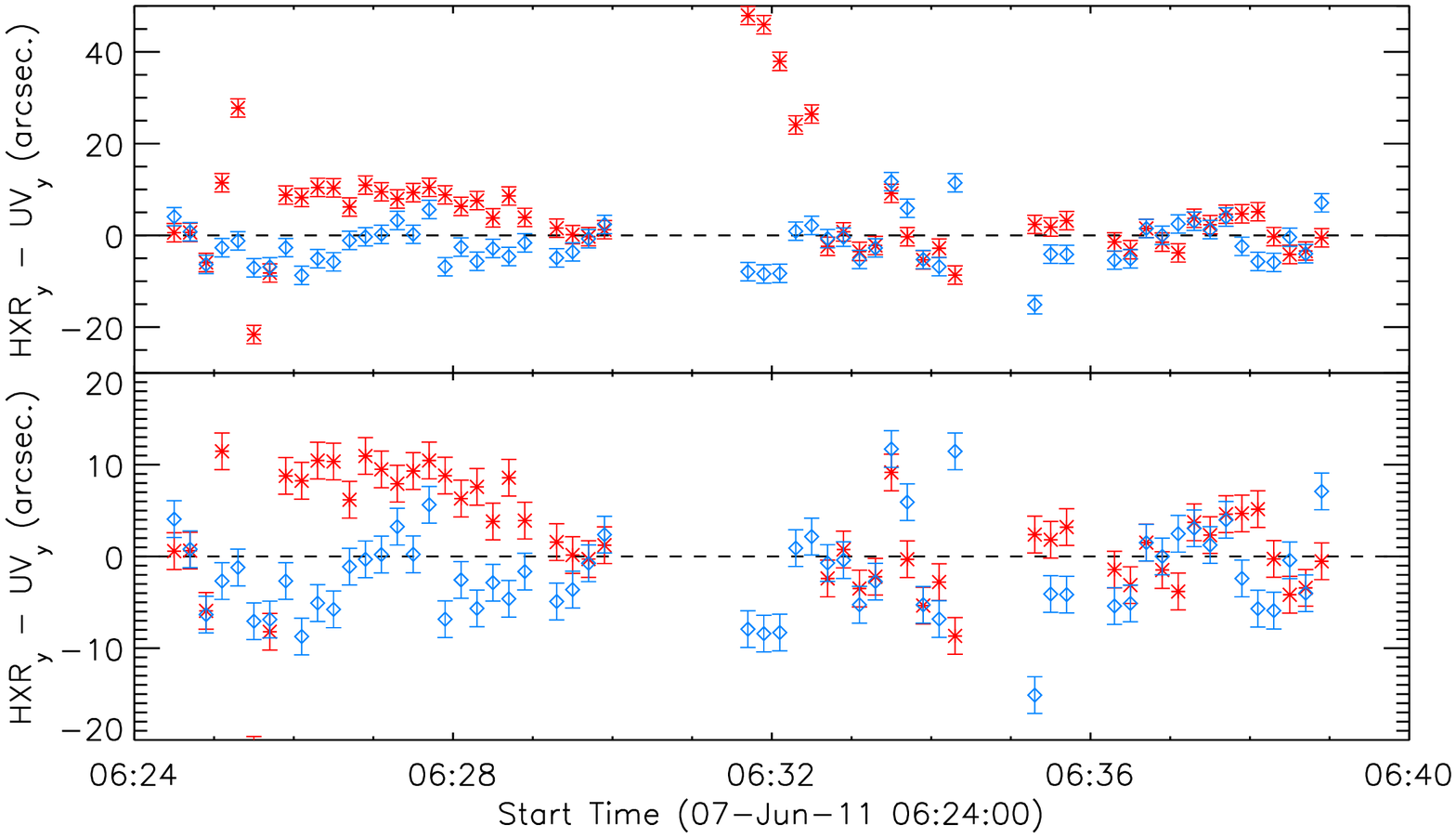}
\caption{Top: Difference between the position of the hard X-ray footpoint sources along the ribbon and the position of the brightest point in the corresponding UV ribbon. The difference between the location of F1 and the left-hand UV ribbon (A) maximum is shown in red (stars), while the difference between F2 and the right-hand UV ribbon (B) maximum is shown in blue (diamonds). Bottom: Same as the top panel, but with a zoomed y-axis.}
\label{diff_between_maxes}
\end{center}
\end{figure}

To further illustrate the offset between the hard X-ray and UV emission, we find the point of maximum UV intensity in each ribbon as a function of time. In Figure \ref{diff_between_maxes} we compare these points with the positions of the hard X-ray footpoints, restricting our attention to the y-direction along each ribbon. The comparison between F1 and the left-hand UV ribbon is shown in red, while F2 and the right-hand UV ribbon are shown in blue.

Between 06:24 - 06:30 the offset between the UV and hard X-ray peak positions is evident in both ribbons, with a larger deviation for ribbon A. Dividing the data into two components is instructive. For this purpose we define the `first phase' as between 06:24 - 06:30 UT, and a `second phase' between 06:32 and 06:40 UT. For each phase, we investigate the distribution of the offsets between the X-ray and EUV emission.

In the `first phase' we find that, for ribbon A, the mean offset $\bar{O}_{A1}$ = 5.1'', with a standard deviation of $\sigma_{A1}$ = 8.6''. Meanwhile, for ribbon B we find the offset $\bar{O}_{B1}$ = -2.4'', with standard deviation $\sigma_{B1}$ = 3.7''.

For the `second phase' we exclude the five outlying points observed at around 06:32 UT in ribbon A. Then, repeating the analysis for the `second phase' we find that $\bar{O}_{A2}$ = -0.07'', with standard deviation $\sigma_{A2}$ = 4.0'', while $\bar{O}_{B2}$ = -1.5'' with standard deviation $\sigma_{B2}$ = 5.9''. Hence, for ribbon A a clear difference between the first and second phases is observed. For ribbon B, the offset also decreases during the second phase, while the standard deviation of the distribution is increased.

This change in relative location may be considered in the context of the behaviour of the hard X-ray and UV lightcurves, which show a substantial dip between 06:30 - 06:32 UT (see Figures \ref{lightcurve} and \ref{time_distance}c), before a second phase of enhanced emission after 06:32 UT. One explanation is that this flare is in fact composed of two distinct elements.

Hence, although there is a temporal correlation between the UV ribbon intensity and the hard X-ray flux during the 2011 June 7 event (see also Figure \ref{correlations}, the spatial relationship between them is less clear. This is consistent with previous observations of UV ribbons and hard X-ray sources \citep[e.g.][]{2006ApJ...640..505A, 2011SSRv..159...19F}, but different from previous observations of ``white light'' emission during flares, which have generally been shown to be closely correlated in space with hard X-rays \citep{2006SoPh..234...79H}. This reinforces the idea that a single mechanism cannot be responsible for the observed emission at these wavelengths.

\subsection{Correlations between parameters}
\label{corrs}

The relationship between the various measurable flare parameters provide clues towards explaining the observed X-ray and UV emission in this flare. For example, \citet{2002SoPh..210..307F} suggested that the hard X-ray flux may be linked to the footpoint behaviour, while \citet{2004ApJ...612..546S} presented evidence that the hard X-ray flux above 25 keV was correlated with the motion of thermal looptop sources during three flares. Also, \citet{2005AdSpR..35.1707K} presented evidence that the hard X-ray flux in flares was correlated with $E_{rec}$, the coronal reconnection rate as estimated from observations of the photospheric magnetic field and the footpoint velocity, a result that has been replicated by a number of additional studies \citep[e.g.][]{2007ApJ...654..665T}. Further examples of investigations into flare parameter correlations include \citet{2006A&A...446..675V, 2009A&A...493..241F, 2009ApJ...693..847L, 2013A&A...552A..87W}. Exploring such correlations in the context of X-ray pulsations also allows us to test the validity of current explanations of QPP.

 The relationship between parameters can be investigated utilising the Spearman rank correlation coefficient. The Spearman coefficient is calculated by assigning ranks $x_i$ and $y_i$ to the values of the variables $X$ and $Y$ (e.g. the lowest value of $X$ would be assigned rank 1, the next-lowest rank 2 etc.). If two values are equal then they are assigned an average rank. The coefficient is then calculated as

\begin{equation}
\rho = \frac{\sum_i (x_i - \bar{x}) (y_i - \bar{y})}{\sqrt{\sum_i (x_i - \bar{x})^2 \sum_i (y_i - \bar{y})^2}   }.
\end{equation}

A feature of the Spearman coefficient is that it is non-parametric; a perfectly linear correlation between two variables $X$ and $Y$ is not required in order to find $\rho$ = 1. Instead, $\rho$ = 1 indicates a monotonic dependence between $X$ and $Y$.

\begin{figure}
\begin{center}
\includegraphics[width=8.5cm]{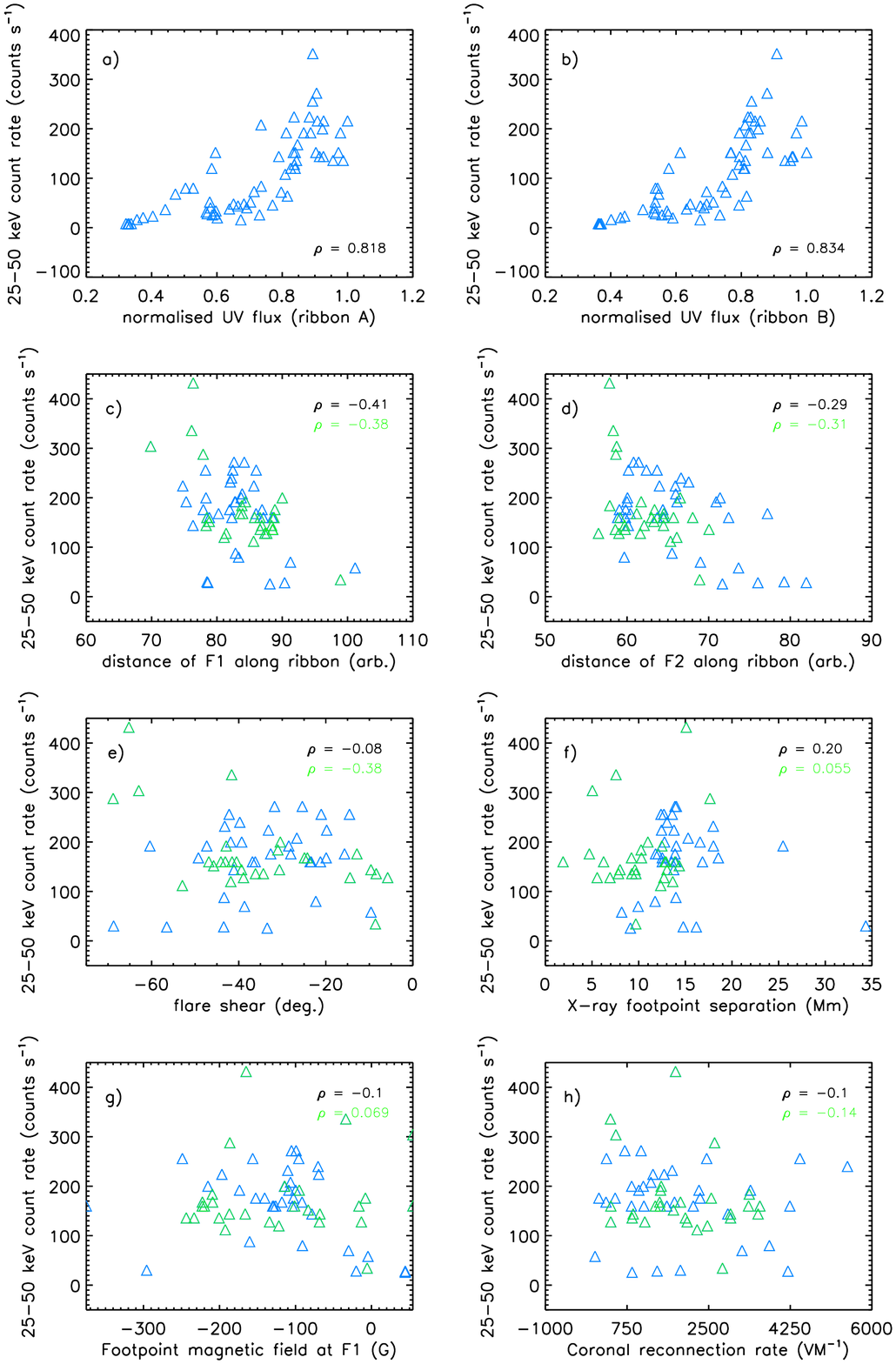}
\caption{Correlations between various measured parameters of the 2011 June 7 flare arcade, as measured between 06:24 and 06:40 UT. a) RHESSI hard X-ray count rates at 25-50 keV versus the UV intensity from ribbon A; b) 25-50 keV X-ray count rates versus UV intensity from ribbon B; c) 25-50 keV X-ray count rates versus position of X-ray footpoint F1; d) 25-50 keV X-ray count rates versus position of X-ray footpoint F2; e) 25- 50 keV X-ray count rates versus estimated flare shear angle; f) 25-50 keV X-ray count rates versus X-ray footpoint separation; g) 25- 50 keV X-ray count rates versus photospheric magnetic field beneath F1 (for brevity, F2 is not shown). h) 25-50 keV X-ray count rates versus estimated coronal reconnection rate $E_{rec}$. The green data denotes measurements made during the `first phase' of the flare, between 06:24 - 06:30 UT. The Spearman rank coefficients are also calculated for these smaller samples, and are marked in green.}
\label{correlations}
\end{center}
\end{figure}

Figure \ref{correlations} shows the correlation between a number of different flare parameters during this event, during the time interval 06:24 - 06:40 UT. Panels a) and b) highlight the similarity of the integrated hard X-ray and UV ribbon emission as a function of time. We find that $\rho$ = 0.82 and $\rho$ = 0.83 for ribbon A and ribbon B respectively, indicating a strong correlation.

The remaining flare arcade parameters are less clearly correlated. For these parameters, we consider both the full set of measurements taken between 06:24 - 06:40 UT, and a subset of these measurements from the `first phase' of the event, encompassing the time interval 06:24 - 06:30 UT. This subset is marked in green in Figure \ref{correlations}. Panels c) and d) test the relationship between the 25-50 keV hard X-ray emission and the motion parallel to the ribbons of footpoints F1 and F2. In both cases, the correlation is weak, with $\rho$ = - 0.41 for F1, and $\rho$ = -0.29 for F2. The result is similar when only the `first phase' is considered. In this case we find instead $\rho$ = -0.38 and $\rho$ = -0.31.

The remaining four panels represent null results: panels e), f), g) and h) show almost no correlation between the hard X-ray count rates and the photospheric magnetic field, $E_{rec}$, the flare shear or the footpoint separation, either in the first phase or over the full event interval. The absence of correlation between the magnetic field data and the X-ray count rates contrasts with previous observations of flares \citep[e.g.][]{2005AdSpR..35.1707K, 2007ApJ...654..665T, 2009SoPh..255..107Q} where a relationship has been found. However, it should not be too surprising that an observable relationship is absent from the 2011 June 7 flare; as pointed out by \citet{2009A&A...493..241F, 2011SSRv..159...19F} the quantity $v_{fp}B_{ph}$ is directly related to the coronal field only in 2-dimensional scenarios. During the flare impulsive phase, the situation is clearly more complex, and the relationship between the photospheric and the coronal magnetic field is unclear. Additionally, the uncertainties associated with studying short time-scale lightcurve features such as QPP make the unmasking of any relationships more difficult.

From Figure \ref{correlations} it can be concluded that the X-ray and UV lightcurves, although strongly correlated with each other - particularly interesting in the context of QPP in hard X-rays - are only weakly dependent on the other observed arcade parameters. This is further complicated by the observation in Section \ref{uv_ribbons} that the hard X-ray and UV brightest points are offset, particularly during the early portion of the event. 

This weak correlation between X-ray count rates and footpoint properties builds on the conclusions of \citet{2012ApJ...748..139I}, who compared the interval between X-ray pulses with the footpoint separation and velocity as a function of time for three flares. No clear link between these observables was found. For the 2011 June 7 event, we also observe QPP in hard X-ray emission that are apparently not associated with any clear spatial pattern (see Figure \ref{dist_from_line} and Figure \ref{correlations}). Furthermore, given the general correlation between the hard X-ray and UV emission during this flare (see Figure \ref{time_distance}), it is possible that QPP are also present in the UV ribbon emission - this possiblity cannot be definitively resolved due to the limited time resolution of SDO/AIA.

The lack of correlation between the temporal and spatial emission suggests that, at least for this flare, QPP are not caused by a secondary effect such as the excitation of waves, and may instead be a fundamental signature of the flare energy release.

\section{Discussion and Conclusions}

We have studied the hard X-ray and UV emission during the impulsive phase of the 2011 June 7 solar flare, utilising data from the RHESSI and SDO/AIA instruments. We have focused on three main features of interest during this event; 1) The presence of QPP in the integrated flare lightcurves, 2) The presence of substantial and unusual hard X-ray footpoint motions, and 3) The presence of two UV ribbons, which also exhibit QPP in their integrated lightcurves. The spatial and temporal behaviour of this emission provides us with clues as to the fundamental nature of flares.

Firstly, we have shown the presence of hard X-ray footpoint motion reversal during the 2011 June 7 flare. The X-ray sources are observed to move parallel to the UV ribbons with mean velocity in the range 30 - 60 km/s. The left-hand footpoint, F1, reverses direction on two occasions. The hard X-ray sources also move perpendicular to the neutral line with a mean separation velocity of ~12 km/s.

To the best of our knowledge, the only previous clear observation of footpoint reversal behaviour was made by \citet{2006ApJ...641.1217C}, who observed similar `zig-zag' motions in a white-light footpoint from the 2002 September 30 flare. Such an observation poses fundamental questions about the nature of flare energy release, particularly due to the simultaneous observation of QPP; during this event there are approximately 10 discernable pulses in the X-ray lightcurves (see Figure \ref{lightcurve}). Further questions are raised by the fact that the hard X-ray source locations associated with the peak times of the pulsations appear to be randomly distributed throughout the flare arcade (see Figure \ref{dist_from_line}a). 

Figure \ref{dist_from_line} presents measurements of additional directly observable properties of the arcade, including the thermal X-ray source motion, magnetic field properties, footpoint separation and flare shear. We can see that, although the locations of the 10 measured pulses are closely associated with the UV ribbons, and tend to be associated with regions of significant photospheric magnetic field, they are not necessarily associated with the regions of \textit{strongest} magnetic field (Figure \ref{dist_from_line}b). Over longer timescales it can be seen that many of the measured properties exhibite general trends. For example, the estimated flare shear angle of the reconnecting loops shows a decreasing trend over the long term, consistent with previous observations that reconnection begins with the most sheared regions of the arcade and progresses to less sheared regions \citep[e.g.][]{2001SoPh..204...55M, 2007ApJ...655..606S, 2009ApJ...693..847L}. Meanwhile, the hard X-ray footpoint separation gradually increases over time, while the ratio of the photospheric field underneath footpoints F1 and F2 shows a trend towards equalisation. However, in each case the behaviour over shorter timescales is fluid. 

A similar discrepancy between short and long timescales is observed in the comparison between the thermal and non-thermal X-ray source motion. In the longer timescale the thermal and non-thermal behaviour is consistent, with the thermal source exibiting smooth motion at least partly associated with height increase, while the hard X-ray sources grow further apart at a rate of $v \approx$ 12 km/s. This is also consistent with the outward expansion of the UV ribbons. However, at shorter timescales, the motion of the hard X-ray sources parallel to the flare ribbon and the reversing motion of F1 does not appear to correspond to any fine structure in the thermal source motion.

The UV ribbons move apart a velocity consistent with the outward motion component of the hard X-ray footpoints during the flare impulsive phase. The integrated emission from these ribbons is strongly correlated with the hard X-ray emission (see Figure \ref{time_distance} and Figure \ref{correlations}). However, the brightest points in the UV emission from each ribbon are not cospatial with the associated brightest hard X-ray footpoints during the early flare times, (see Figures \ref{time_distance},\ref{diff_between_maxes}) with offsets of 5 - 10'' between 06:24 - 06:30 UT . Beyond 06:32 UT the locations of the hard X-ray and UV brightest areas become approximately cospatial. Similar to the contrast between the X-ray thermal and non-thermal motions, the UV and hard X-ray behaviour is consistent over longer timescales but not over short timescales. The rapid motion of the X-ray footpoints along the flare ribbons has no noticeable UV counterpart. Instead, the brightest areas in the UV emission remain approximately static as the flare progresses, except for their gradual outward expansion.  

In order to understand whether there is a relationship between the short-timescale behaviour of the measured flare properties and the hard X-ray and UV lightcurves, we carried out correlation studies as detailed in Section \ref{corrs}. It was shown that the QPP observed in hard X-ray and UV emission is only weakly correlated with the estimated shear angle, hard X-ray footpoint location, and footpoint separation, and uncorrelated with the estimated magnetic field properties underneath the X-ray footpoints. This lack of correlation is in contrast to the results obtained in some previous studies of footpoint motion and magnetic field properties \citep[e.g.][]{2005AdSpR..35.1707K, 2007ApJ...654..665T}. 

One possible interpretation of the hard X-ray motions and pulsations in this event is that magnetohydrodynamic waves excited in the flare arcade by the main energy release drive the short timescale behaviour. For example, \citet{2011ApJ...730L..27N} recently proposed that slow waves are capable of propagating and reflecting along flare arcades, which could in principle explain both footpoint motion and the generation of X-ray pulsations \citep[see also][]{2011A&A...536A..68G}. However, if QPP were indeed generated as a secondary effect, such as via propagating MHD waves, then greater correlations between the measured flare parameters might be expected - for example, in the \citet{2011ApJ...730L..27N} model the characteristic timescale of lightcurve pulses should be a function of the footpoint or ribbon separation. Additionally, it can be surmised from Figure \ref{dist_from_line}a and Figure \ref{dist_from_line}b that the spatial distribution of the X-ray pulses does not follow an obvious pattern along the arcade, as might be expected from such a model. A further limitation is that, in order to account for the 2011 June 7 observation, it would have to be shown that these slow waves can also reflect and reverse direction along a flare arcade.

Alternatively, X-ray pulses and footpoints can be considered as a direct signature of quasi-oscillatory or `bursty' reconnection \citep[e.g.][]{2006ApJ...642.1177L, 2009A&A...494..329M,2011ApJ...730...90G, 2012A&A...548A..98M, 2012ApJ...749...30M}. In this scenario, the location, extent and motion of the reconnection site(s) remains unclear, although it might be expected that the observed QPP would be related to the estimated coronal reconnection rate $E_{rec}$. However, estimation of this quantity based on the observed footpoint motion and photospheric field may be too great an oversimplification during a flare's impulsive phase \citep{2011SSRv..159...19F}. For the 2011 June 7 event, the overall motion of the thermal X-ray source shows possible motion both parallel to the arcade and increasing in height, but shows little evidence of fine structure that might correlate with the hard X-ray source motions. Meanwhile the observed hard X-ray footpoint reversals imply movement of particle acceleration regions along the arcade or the presence of multiple, discrete reconnection sites. This is further complicated by the spatial offset of the brightest points in the UV ribbons from the brightest hard X-ray footpoints, particlarly early in the impulsive phase. Regardless, there is no obvious explanation for the observed reversal of hard X-ray footpoints.

The lack of correlation between the short-timescale footpoint motion and the other observed arcade parameters, in contrast with the strong correlation between the X-ray and UV lightcurves, leads us to conclude that the true driver of QPP in this flare is an as-yet unobserved property of the flare reconnection region. The observed arcade parameters, such as the locations of hard X-ray sources and the photospheric field underneath the footpoints, are secondary considerations which are broadly independent of the spatially integrated emission during QPP. By contrast, the long-term behaviour of both the X-ray and UV emission is explained via a gradual expansion of the arcade and increase in height of the reconnection region, consistent with the standard flare model. These observations present a challenge for current models of flare energy release.

\begin{acknowledgements}
ARI is grateful to Dr Brian Dennis, Dr Gordon Holman, Dr Joel Allred, Dr Adam Kowalski and Dr Lyndsay Fletcher for helpful discussions which helped to improve this paper, as well as to the RHESSI and SDO science teams for making available the data used in this study.
\end{acknowledgements}

\bibliographystyle{apj}
\bibliography{20110607_qpp}

\begin{thebibliography}{71}
\expandafter\ifx\csname natexlab\endcsname\relax\def\natexlab#1{#1}\fi

\bibitem[{{Alexander} \& {Coyner}(2006)}]{2006ApJ...640..505A}
{Alexander}, D., \& {Coyner}, A.~J. 2006, \apj, 640, 505

\bibitem[{{Asai} {et~al.}(2001){Asai}, {Shimojo}, {Isobe}, {Morimoto},
  {Yokoyama}, {Shibasaki}, \& {Nakajima}}]{2001ApJ...562L.103A}
{Asai}, A., {Shimojo}, M., {Isobe}, H., {et~al.} 2001, \apjl, 562, L103

\bibitem[{{Bogachev} {et~al.}(2005){Bogachev}, {Somov}, {Kosugi}, \&
  {Sakao}}]{2005ApJ...630..561B}
{Bogachev}, S.~A., {Somov}, B.~V., {Kosugi}, T., \& {Sakao}, T. 2005, \apj,
  630, 561

\bibitem[{{Brown}(1971)}]{1971SoPh...18..489B}
{Brown}, J.~C. 1971, \solphys, 18, 489

\bibitem[{{Carmichael}(1964)}]{1964NASSP..50..451C}
{Carmichael}, H. 1964, NASA Special Publication, 50, 451

\bibitem[{{Chen} \& {Ding}(2006)}]{2006ApJ...641.1217C}
{Chen}, Q.~R., \& {Ding}, M.~D. 2006, \apj, 641, 1217

\bibitem[{{Chiu}(1970)}]{1970SoPh...13..420C}
{Chiu}, Y.~T. 1970, \solphys, 13, 420

\bibitem[{{Demoulin} {et~al.}(1997){Demoulin}, {Bagala}, {Mandrini}, {Henoux},
  \& {Rovira}}]{1997A&A...325..305D}
{Demoulin}, P., {Bagala}, L.~G., {Mandrini}, C.~H., {Henoux}, J.~C., \&
  {Rovira}, M.~G. 1997, \aap, 325, 305

\bibitem[{{Dennis} \& {Pernak}(2009)}]{2009ApJ...698.2131D}
{Dennis}, B.~R., \& {Pernak}, R.~L. 2009, \apj, 698, 2131

\bibitem[{{Ding} {et~al.}(2003){Ding}, {Liu}, {Yeh}, \&
  {Li}}]{2003A&A...403.1151D}
{Ding}, M.~D., {Liu}, Y., {Yeh}, C.-T., \& {Li}, J.~P. 2003, \aap, 403, 1151

\bibitem[{{Dolla} {et~al.}(2012){Dolla}, {Marqu{\'e}}, {Seaton}, {Van
  Doorsselaere}, {Dominique}, {Berghmans}, {Cabanas}, {De Groof}, {Schmutz},
  {Verdini}, {West}, {Zender}, \& {Zhukov}}]{2012ApJ...749L..16D}
{Dolla}, L., {Marqu{\'e}}, C., {Seaton}, D.~B., {et~al.} 2012, \apjl, 749, L16

\bibitem[{{Fermi-LAT collaboration}(2013)}]{2013arXiv1304.3749F}
{Fermi-LAT collaboration}. 2013, ArXiv e-prints

\bibitem[{{Fletcher}(2009)}]{2009A&A...493..241F}
{Fletcher}, L. 2009, \aap, 493, 241

\bibitem[{{Fletcher} \& {Hudson}(2002)}]{2002SoPh..210..307F}
{Fletcher}, L., \& {Hudson}, H.~S. 2002, \solphys, 210, 307

\bibitem[{{Fletcher} {et~al.}(2004){Fletcher}, {Pollock}, \&
  {Potts}}]{2004SoPh..222..279F}
{Fletcher}, L., {Pollock}, J.~A., \& {Potts}, H.~E. 2004, \solphys, 222, 279

\bibitem[{{Fletcher} {et~al.}(2011){Fletcher}, {Dennis}, {Hudson}, {Krucker},
  {Phillips}, {Veronig}, {Battaglia}, {Bone}, {Caspi}, {Chen}, {Gallagher},
  {Grigis}, {Ji}, {Liu}, {Milligan}, \& {Temmer}}]{2011SSRv..159...19F}
{Fletcher}, L., {Dennis}, B.~R., {Hudson}, H.~S., {et~al.} 2011, \ssr, 159, 19

\bibitem[{{Foullon} {et~al.}(2005){Foullon}, {Verwichte}, {Nakariakov}, \&
  {Fletcher}}]{2005A&A...440L..59F}
{Foullon}, C., {Verwichte}, E., {Nakariakov}, V.~M., \& {Fletcher}, L. 2005,
  \aap, 440, L59

\bibitem[{{Gilbert} {et~al.}(2013){Gilbert}, {Inglis}, {Mays}, {Ofman},
  {Thompson}, \& {Young}}]{Gilbert2013}
{Gilbert}, H.~R., {Inglis}, A.~R., {Mays}, M.~L., {et~al.} 2013, ApJL,
  submitted

\bibitem[{{Grechnev} {et~al.}(2003){Grechnev}, {White}, \&
  {Kundu}}]{2003ApJ...588.1163G}
{Grechnev}, V.~V., {White}, S.~M., \& {Kundu}, M.~R. 2003, \apj, 588, 1163

\bibitem[{{Grigis} \& {Benz}(2005)}]{2005ApJ...625L.143G}
{Grigis}, P.~C., \& {Benz}, A.~O. 2005, \apjl, 625, L143

\bibitem[{{Gruszecki} \& {Nakariakov}(2011)}]{2011A&A...536A..68G}
{Gruszecki}, M., \& {Nakariakov}, V.~M. 2011, \aap, 536, A68

\bibitem[{{Guidoni} \& {Longcope}(2011)}]{2011ApJ...730...90G}
{Guidoni}, S.~E., \& {Longcope}, D.~W. 2011, \apj, 730, 90

\bibitem[{{Hirayama}(1974)}]{1974SoPh...34..323H}
{Hirayama}, T. 1974, \solphys, 34, 323

\bibitem[{{Holman} {et~al.}(2011){Holman}, {Aschwanden}, {Aurass}, {Battaglia},
  {Grigis}, {Kontar}, {Liu}, {Saint-Hilaire}, \&
  {Zharkova}}]{2011SSRv..159..107H}
{Holman}, G.~D., {Aschwanden}, M.~J., {Aurass}, H., {et~al.} 2011, \ssr, 159,
  107

\bibitem[{{Hudson}(1972)}]{1972SoPh...24..414H}
{Hudson}, H.~S. 1972, \solphys, 24, 414

\bibitem[{{Hudson} {et~al.}(2006){Hudson}, {Wolfson}, \&
  {Metcalf}}]{2006SoPh..234...79H}
{Hudson}, H.~S., {Wolfson}, C.~J., \& {Metcalf}, T.~R. 2006, \solphys, 234, 79

\bibitem[{{Hurford} {et~al.}(2002){Hurford}, {Schmahl}, {Schwartz}, {Conway},
  {Aschwanden}, {Csillaghy}, {Dennis}, {Johns-Krull}, {Krucker}, {Lin},
  {McTiernan}, {Metcalf}, {Sato}, \& {Smith}}]{2002SoPh..210...61H}
{Hurford}, G.~J., {Schmahl}, E.~J., {Schwartz}, R.~A., {et~al.} 2002, \solphys,
  210, 61

\bibitem[{{Inglis} \& {Dennis}(2012)}]{2012ApJ...748..139I}
{Inglis}, A.~R., \& {Dennis}, B.~R. 2012, \apj, 748, 139

\bibitem[{{Inglis} {et~al.}(2008){Inglis}, {Nakariakov}, \&
  {Melnikov}}]{2008A&A...487.1147I}
{Inglis}, A.~R., {Nakariakov}, V.~M., \& {Melnikov}, V.~F. 2008, \aap, 487,
  1147

\bibitem[{{Innes} {et~al.}(2012){Innes}, {Cameron}, {Fletcher}, {Inhester}, \&
  {Solanki}}]{2012A&A...540L..10I}
{Innes}, D.~E., {Cameron}, R.~H., {Fletcher}, L., {Inhester}, B., \& {Solanki},
  S.~K. 2012, \aap, 540, L10

\bibitem[{{Ji} {et~al.}(2007){Ji}, {Huang}, \& {Wang}}]{2007ApJ...660..893J}
{Ji}, H., {Huang}, G., \& {Wang}, H. 2007, \apj, 660, 893

\bibitem[{{Kopp} \& {Pneuman}(1976)}]{1976SoPh...50...85K}
{Kopp}, R.~A., \& {Pneuman}, G.~W. 1976, \solphys, 50, 85

\bibitem[{{Krucker} {et~al.}(2005){Krucker}, {Fivian}, \&
  {Lin}}]{2005AdSpR..35.1707K}
{Krucker}, S., {Fivian}, M.~D., \& {Lin}, R.~P. 2005, Advances in Space
  Research, 35, 1707

\bibitem[{{Krucker} {et~al.}(2003){Krucker}, {Hurford}, \&
  {Lin}}]{2003ApJ...595L.103K}
{Krucker}, S., {Hurford}, G.~J., \& {Lin}, R.~P. 2003, \apjl, 595, L103

\bibitem[{{Lee} \& {Gary}(2008)}]{2008ApJ...685L..87L}
{Lee}, J., \& {Gary}, D.~E. 2008, \apjl, 685, L87

\bibitem[{{Leibacher} {et~al.}(2010){Leibacher}, {Sakurai}, {Schrijver}, \&
  {van Driel-Gesztelyi}}]{2010SoPh..263....1L}
{Leibacher}, J., {Sakurai}, T., {Schrijver}, C., \& {van Driel-Gesztelyi}.
  2010, \solphys, 263, 1

\bibitem[{{Li} {et~al.}(2012){Li}, {Zhang}, {Yang}, \&
  {Liu}}]{2012ApJ...746...13L}
{Li}, T., {Zhang}, J., {Yang}, S., \& {Liu}, W. 2012, \apj, 746, 13

\bibitem[{{Li} \& {Gan}(2008)}]{2008SoPh..247...77L}
{Li}, Y.~P., \& {Gan}, W.~Q. 2008, \solphys, 247, 77

\bibitem[{{Lin} {et~al.}(2002){Lin}, {Dennis}, {Hurford}, {Smith}, {Zehnder},
  {Harvey}, {Curtis}, {Pankow}, {Turin}, {Bester}, {Csillaghy}, {Lewis},
  {Madden}, {van Beek}, {Appleby}, {Raudorf}, {McTiernan}, {Ramaty}, {Schmahl},
  {Schwartz}, {Krucker}, {Abiad}, {Quinn}, {Berg}, {Hashii}, {Sterling},
  {Jackson}, {Pratt}, {Campbell}, {Malone}, {Landis}, {Barrington-Leigh},
  {Slassi-Sennou}, {Cork}, {Clark}, {Amato}, {Orwig}, {Boyle}, {Banks},
  {Shirey}, {Tolbert}, {Zarro}, {Snow}, {Thomsen}, {Henneck}, {McHedlishvili},
  {Ming}, {Fivian}, {Jordan}, {Wanner}, {Crubb}, {Preble}, {Matranga}, {Benz},
  {Hudson}, {Canfield}, {Holman}, {Crannell}, {Kosugi}, {Emslie}, {Vilmer},
  {Brown}, {Johns-Krull}, {Aschwanden}, {Metcalf}, \& {Conway}}]{Lin2002}
{Lin}, R.~P., {Dennis}, B.~R., {Hurford}, G.~J., {et~al.} 2002, \solphys, 210,
  3

\bibitem[{{Linton} \& {Longcope}(2006)}]{2006ApJ...642.1177L}
{Linton}, M.~G., \& {Longcope}, D.~W. 2006, \apj, 642, 1177

\bibitem[{{Liu} {et~al.}(2009){Liu}, {Petrosian}, {Dennis}, \&
  {Holman}}]{2009ApJ...693..847L}
{Liu}, W., {Petrosian}, V., {Dennis}, B.~R., \& {Holman}, G.~D. 2009, \apj,
  693, 847

\bibitem[{{Masuda} {et~al.}(2001){Masuda}, {Kosugi}, \&
  {Hudson}}]{2001SoPh..204...55M}
{Masuda}, S., {Kosugi}, T., \& {Hudson}, H.~S. 2001, \solphys, 204, 55

\bibitem[{{McLaughlin} {et~al.}(2012{\natexlab{a}}){McLaughlin}, {Thurgood}, \&
  {MacTaggart}}]{2012A&A...548A..98M}
{McLaughlin}, J.~A., {Thurgood}, J.~O., \& {MacTaggart}, D. 2012{\natexlab{a}},
  \aap, 548, A98

\bibitem[{{McLaughlin} {et~al.}(2012{\natexlab{b}}){McLaughlin}, {Verth},
  {Fedun}, \& {Erd{\'e}lyi}}]{2012ApJ...749...30M}
{McLaughlin}, J.~A., {Verth}, G., {Fedun}, V., \& {Erd{\'e}lyi}, R.
  2012{\natexlab{b}}, \apj, 749, 30

\bibitem[{{Meegan} {et~al.}(2009){Meegan}, {Lichti}, {Bhat}, {Bissaldi},
  {Briggs}, {Connaughton}, {Diehl}, {Fishman}, {Greiner}, {Hoover}, {van der
  Horst}, {von Kienlin}, {Kippen}, {Kouveliotou}, {McBreen}, {Paciesas},
  {Preece}, {Steinle}, {Wallace}, {Wilson}, \&
  {Wilson-Hodge}}]{2009ApJ...702..791M}
{Meegan}, C., {Lichti}, G., {Bhat}, P.~N., {et~al.} 2009, \apj, 702, 791

\bibitem[{{Melnikov} {et~al.}(2005){Melnikov}, {Reznikova}, {Shibasaki}, \&
  {Nakariakov}}]{2005A&A...439..727M}
{Melnikov}, V.~F., {Reznikova}, V.~E., {Shibasaki}, K., \& {Nakariakov}, V.~M.
  2005, \aap, 439, 727

\bibitem[{{Metcalf} {et~al.}(1990){Metcalf}, {Canfield}, {Avrett}, \&
  {Metcalf}}]{1990ApJ...350..463M}
{Metcalf}, T.~R., {Canfield}, R.~C., {Avrett}, E.~H., \& {Metcalf}, F.~T. 1990,
  \apj, 350, 463

\bibitem[{{Murray} {et~al.}(2009){Murray}, {van Driel-Gesztelyi}, \&
  {Baker}}]{2009A&A...494..329M}
{Murray}, M.~J., {van Driel-Gesztelyi}, L., \& {Baker}, D. 2009, \aap, 494, 329

\bibitem[{{Najita} \& {Orrall}(1970)}]{1970SoPh...15..176N}
{Najita}, K., \& {Orrall}, F.~Q. 1970, \solphys, 15, 176

\bibitem[{{Nakariakov} {et~al.}(2010){Nakariakov}, {Foullon}, {Myagkova}, \&
  {Inglis}}]{2010ApJ...708L..47N}
{Nakariakov}, V.~M., {Foullon}, C., {Myagkova}, I.~N., \& {Inglis}, A.~R. 2010,
  \apjl, 708, L47

\bibitem[{{Nakariakov} \& {Melnikov}(2009)}]{2009SSRv..149..119N}
{Nakariakov}, V.~M., \& {Melnikov}, V.~F. 2009, \ssr, 149, 119

\bibitem[{{Nakariakov} \& {Zimovets}(2011)}]{2011ApJ...730L..27N}
{Nakariakov}, V.~M., \& {Zimovets}, I.~V. 2011, \apjl, 730, L27

\bibitem[{{Parks} \& {Winckler}(1969)}]{1969ApJ...155L.117P}
{Parks}, G.~K., \& {Winckler}, J.~R. 1969, \apjl, 155, L117

\bibitem[{{Pesnell} {et~al.}(2012){Pesnell}, {Thompson}, \&
  {Chamberlin}}]{2012SoPh..275....3P}
{Pesnell}, W.~D., {Thompson}, B.~J., \& {Chamberlin}, P.~C. 2012, \solphys,
  275, 3

\bibitem[{{Qiu} {et~al.}(2009){Qiu}, {Gary}, \&
  {Fleishman}}]{2009SoPh..255..107Q}
{Qiu}, J., {Gary}, D.~E., \& {Fleishman}, G.~D. 2009, \solphys, 255, 107

\bibitem[{{Qiu} {et~al.}(2010){Qiu}, {Liu}, {Hill}, \&
  {Kazachenko}}]{2010ApJ...725..319Q}
{Qiu}, J., {Liu}, W., {Hill}, N., \& {Kazachenko}, M. 2010, \apj, 725, 319

\bibitem[{{Qiu} {et~al.}(2013){Qiu}, {Sturrock}, {Longcope}, {James},
  {Klimchuk}, {Wen-Juan}, \& {Liu}}]{2013arXiv1305.6899Q}
{Qiu}, J., {Sturrock}, Z., {Longcope}, D.~W., {et~al.} 2013, ArXiv e-prints

\bibitem[{{Qiu} {et~al.}(2004){Qiu}, {Wang}, {Cheng}, \&
  {Gary}}]{2004ApJ...604..900Q}
{Qiu}, J., {Wang}, H., {Cheng}, C.~Z., \& {Gary}, D.~E. 2004, \apj, 604, 900

\bibitem[{{Reznikova} \& {Shibasaki}(2011)}]{2011A&A...525A.112R}
{Reznikova}, V.~E., \& {Shibasaki}, K. 2011, \aap, 525, A112

\bibitem[{{Saba} {et~al.}(2006){Saba}, {Gaeng}, \&
  {Tarbell}}]{2006ApJ...641.1197S}
{Saba}, J.~L.~R., {Gaeng}, T., \& {Tarbell}, T.~D. 2006, \apj, 641, 1197

\bibitem[{{Sturrock}(1966)}]{1966Natur.211..695S}
{Sturrock}, P.~A. 1966, \nat, 211, 695

\bibitem[{{Su} {et~al.}(2007){Su}, {Golub}, \& {Van
  Ballegooijen}}]{2007ApJ...655..606S}
{Su}, Y., {Golub}, L., \& {Van Ballegooijen}, A.~A. 2007, \apj, 655, 606

\bibitem[{{Sui} {et~al.}(2004){Sui}, {Holman}, \&
  {Dennis}}]{2004ApJ...612..546S}
{Sui}, L., {Holman}, G.~D., \& {Dennis}, B.~R. 2004, \apj, 612, 546

\bibitem[{{Temmer} {et~al.}(2007){Temmer}, {Veronig}, {Vr{\v s}nak}, \&
  {Miklenic}}]{2007ApJ...654..665T}
{Temmer}, M., {Veronig}, A.~M., {Vr{\v s}nak}, B., \& {Miklenic}, C. 2007,
  \apj, 654, 665

\bibitem[{{Veronig} {et~al.}(2006){Veronig}, {Karlick{\'y}}, {Vr{\v s}nak},
  {Temmer}, {Magdaleni{\'c}}, {Dennis}, {Otruba}, \&
  {P{\"o}tzi}}]{2006A&A...446..675V}
{Veronig}, A.~M., {Karlick{\'y}}, M., {Vr{\v s}nak}, B., {et~al.} 2006, \aap,
  446, 675

\bibitem[{{Warmuth} \& {Mann}(2013)}]{2013A&A...552A..87W}
{Warmuth}, A., \& {Mann}, G. 2013, \aap, 552, A87

\bibitem[{{Williams} {et~al.}(2013){Williams}, {Baker}, \& {van
  Driel-Gesztelyi}}]{2013ApJ...764..165W}
{Williams}, D.~R., {Baker}, D., \& {van Driel-Gesztelyi}, L. 2013, \apj, 764,
  165

\bibitem[{{Woods} {et~al.}(2012){Woods}, {Eparvier}, {Hock}, {Jones},
  {Woodraska}, {Judge}, {Didkovsky}, {Lean}, {Mariska}, {Warren}, {McMullin},
  {Chamberlin}, {Berthiaume}, {Bailey}, {Fuller-Rowell}, {Sojka}, {Tobiska}, \&
  {Viereck}}]{2012SoPh..275..115W}
{Woods}, T.~N., {Eparvier}, F.~G., {Hock}, R., {et~al.} 2012, \solphys, 275,
  115

\bibitem[{{Xu} {et~al.}(2006){Xu}, {Cao}, {Liu}, {Yang}, {Jing}, {Denker},
  {Emslie}, \& {Wang}}]{2006ApJ...641.1210X}
{Xu}, Y., {Cao}, W., {Liu}, C., {et~al.} 2006, \apj, 641, 1210

\bibitem[{{Yang} {et~al.}(2009){Yang}, {Cheng}, {Krucker}, {Lin}, \&
  {Ip}}]{2009ApJ...693..132Y}
{Yang}, Y.-H., {Cheng}, C.~Z., {Krucker}, S., {Lin}, R.~P., \& {Ip}, W.~H.
  2009, \apj, 693, 132

\bibitem[{{Zimovets} \& {Struminsky}(2009)}]{2009SoPh..258...69Z}
{Zimovets}, I.~V., \& {Struminsky}, A.~B. 2009, \solphys, 258, 69

\end{thebibliography}

\end{document}